\documentclass[11pt]{article}
\usepackage{newpasp}
\usepackage{psfig}
\usepackage{epsf}
\begin{document}
\title{Chandra Observations of Three SDSS Quasars at z$\approx$6}

\author{D. A. Schwartz}
\affil{Harvard-Smithsonian Center for Astrophysics, 60 Garden St.,
Cambridge MA 02138}
\author{C. C. Cheung, J. F. C. Wardle}
\affil{Physics Department, Brandeis University, Waltham, MA 02454}

%\begin{abstract}

%\end{abstract}

\section{Observations}
The Sloan Digital Sky Survey (SDSS) quasars at redshift z $\approx$ 6
(Fan et al. 2001) are the most distant condensed objects known in the
universe. X-ray observations of them are of great importance; e.g., to
study the cosmic time evolution of X-ray emission and to assess the
detectability of quasar X-ray emission at even larger redshifts.  All
three quasars were easily detected in X-rays even in these relatively
short observations (Table 1). However, as shown in Schwartz (2002c),
\emph{none were totally isolated point sources.}

\begin{table}[h]
    %>>>> [h] means place table here

\caption{\label{tab:sdssObs}\emph{Chandra} Observations of the SDSS quasars at redshift 6 }
\footnotesize
\scriptsize
%\tablewidth{0pt}
\begin{tabular}{ccccccccc}
\hline
%\tablecolumns{9}

Name$^{a}$ &  &Time  & Core &
& Core &Jet &  & Remarks   \\*
SDSSp  & z$^{a}$ &
ksec & Cnts &L$_{\mathrm{core}}^{b}$
& $\alpha_{\mathrm{ox}}$&

Cnts &L$_{\mathrm{jet}}^{b}$ &
 \\

J083643.85+005453.3  &5.82 &5.686 &21 &2.3&1.66
& $<$ 6.3 & $<$0.70 & Nearby Source \\
J103027.10+052455.0 &6.28  &7.942&6 & 0.55&1.79 &$<$6.3 &$<$0.59  &
Core not a Point\\
J130608.26+035626.3 &5.99  &8.160& 16 & 1.3&1.65 & 7\tablenotemark{c}
&0.57\tablenotemark{c}  & Nearby Jet \\ 

\end{tabular}

$^{a}${Fan et al. 2001}
$^{b}${Rest frame 2 --10 keV luminosity in units of 10$^{45}$ergs
s$^{-1}$. We use H$_{0}$=65, $\Omega_{0}$=0.3,
and $\Omega_{\Lambda}$=0.7}
$^{c}${Significant detection, but jet identification not certain}
\end{table}

We specifically searched for an X-ray jet in these quasars,
following the recognition (Schwartz 2002a,b) that \emph{if} X-ray
emission observed from radio jets at modest redshift is due to inverse
Compton (IC) scattering on the Cosmic Microwave Background (CMB), then similar
jets will have essentially the same surface brightness at arbitrarily
larger redshifts, and may serve as ``beacons'' from the distant
universe.  We find an extended feature 23$\arcsec$ from the quasar
SDSS 1306+0356, (Fig. 1, left) which may be such a jet.  We can
explain the emission as IC/CMB, assuming that the intrinsic properties of
the system are similar to other X-ray jets (Fig 1, right).

In May 2002 we carried out a VLA target-of-opportunity observation of
SDSS 1306 at 1.4 GHz. A two hour
observation in the A configuration gave upper limits to either the
quasar core or the jet of 0.1 mJy (3$\sigma$).  A recent paper by Ivanov (2002)
found a 23rd magnitude galaxy at the location of the jet feature. Both
of the above mitigate against the extended source CXOU
J130609.1+035643.5 being a jet associated with the quasar. A deep,
120 ksec, \emph{Chandra} observation has been approved for cycle 4 to
confirm or refute the jet identification.

\begin{figure}[t]
\plottwo{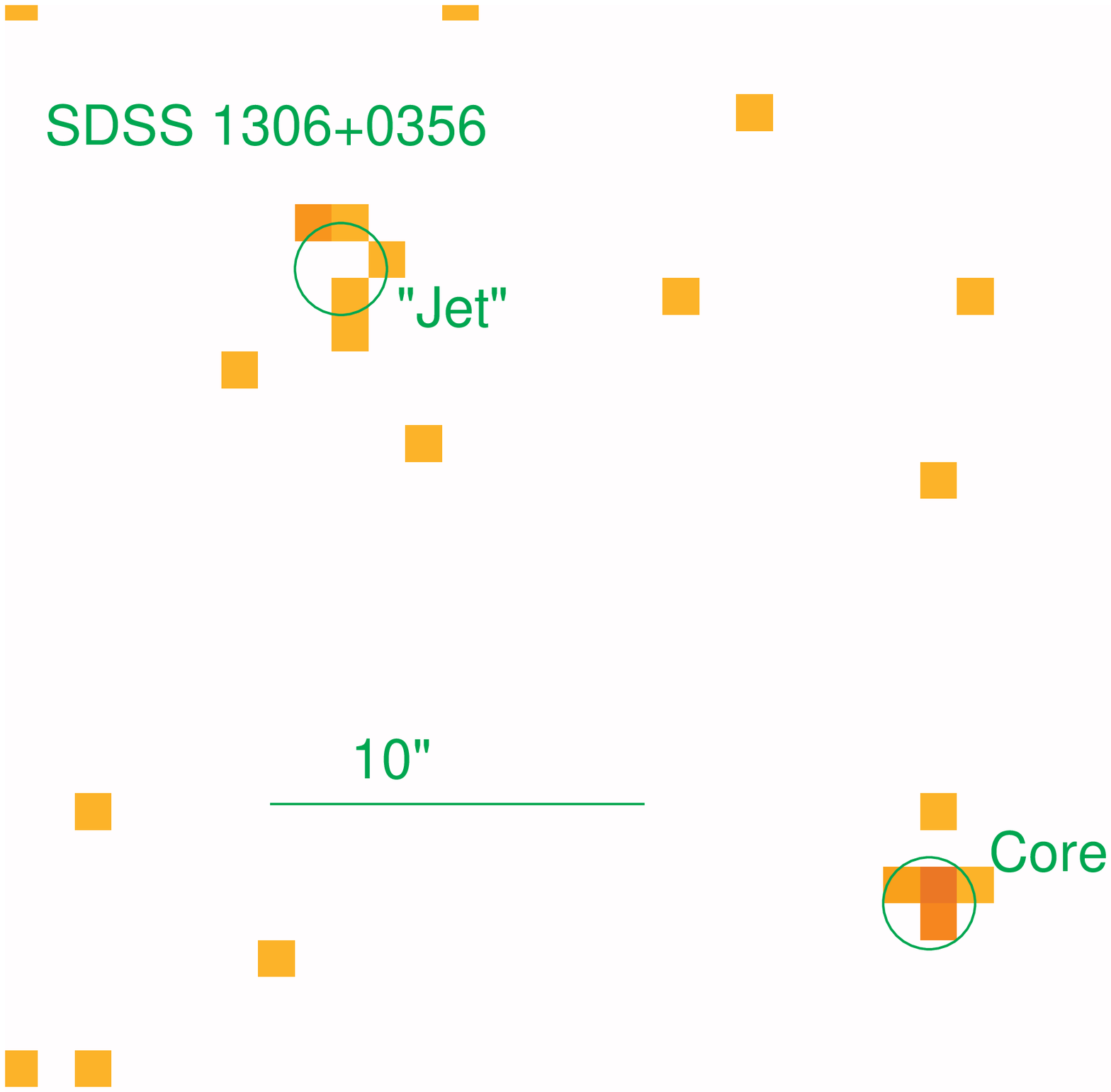}{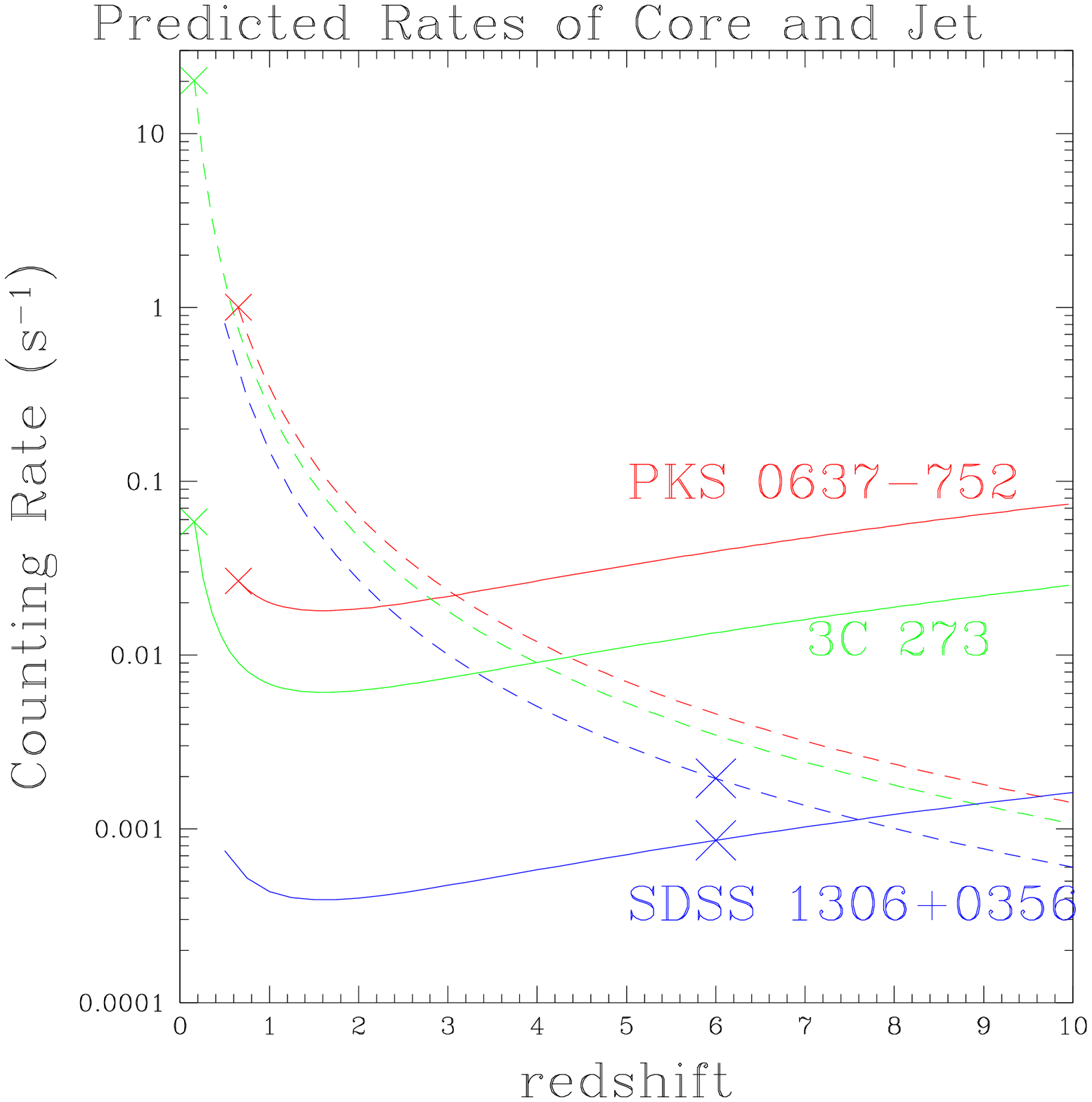}

\caption{(Left) The 0.5 to 7 keV X-rays from SDSS 1306 binned by
1$^{\prime\prime}$. The circle is the 95\% encircled energy diameter,
so the feature we call a jet cannot be a single point source. (Right)
The counting rates of IC/CMB jets, (solid lines) would remain roughly
constant at any redshift, changing only with the small solid angle
change and K-correction. The crosses plot the observed values at the
actual redshifts. The cores, (dashed lines) change as
D$_{\rm{lum}}^{-2}$, so that at redshift z=1, SDSS~1306 would actually
have the lowest jet to core ratio of these three objects.}

\end{figure}

\section{Conclusions}

%\section{CONCLUSIONS}
Note that if the flux of these quasars had been measured within a
60\arcsec\ circle, as commonly done for low redshift quasars using
\emph{ROSAT}, then their $\alpha_{\mathrm{ox}}$ values would seem
smaller (i.e., flatter optical to X-ray slope) by $\approx$0.15 due to nearby contaminating
sources. Both the point-like source SW of SDSS 0836, and the extended
core of SDSS 1030 (Schwartz 2002c, Fig 2), should also be considered as
possible X-ray jets.  Observations of quasars with \emph{Chandra} will
be important to obtain spectral information uncontaminated by nearby
sources and to search for jets and gravitational lensing.

\acknowledgements This research was sponsored in part by NASA contract
NAS8-39073 to the Chandra X-ray Center, and by NSF grant AST 99-00723
to Brandeis. We thank the VLA scheduling committee for their quick
response to our ToO request.

\end{document}